# A feasible roadmap to identifying significant intercellular genomic heterogeneity in deep sequencing data

Guoqiang Yu[1], Niya Wang[1], Roger R. Wang[2], Sean S. Wang[3], Yue Wang[1]

[1]Department of Electrical and Computer Engineering, Virginia Polytechnic Institute and State University, Arlington, VA 22203, USA; [2]Department of Electrical Engineering and Computer Science, University of Michigan, Ann Arbor, MI 48109, USA; [3]Department of Electrical and Computer Engineering, University of Maryland, College Park, MD 20742, USA

*Abstract*—Intercellular heterogeneity serves as both a confounding factor in studying individual clones and an information source in characterizing any heterogeneous tissues, such as blood, tumor systems. Due to inevitable sequencing errors and other sample preparation artifacts such as PCR errors, systematic efforts to characterize intercellular genomic heterogeneity must effectively distinguish genuine clonal sequences from fake derivatives. We developed a novel approach (SIGH) for identifying significant genuine clonal sequences directly from mixed sequencing reads that can improve genomic analyses in many biological contexts. This method offers several attractive features: (1) it automatically estimates the error rate from raw sequence reads and identifies genuine clonal sequences; (2) it is robust to the large variety of error rate due to the various experimental conditions; (3) it is supported by a well-grounded statistical framework that exploits probabilistic characteristics of sequencing errors; (4) its unbiased strategy allows detecting rare clone(s) despite that clone's relative abundance; and (5) it estimates constituent proportions in each sample. Extensive realistic simulation studies show that our method can reliably estimate the error rates and faithfully distinguish the genuine clones from fake derivatives, paving the way for follow-up analysis that is otherwise ruined by the often dominant fake clones.

## I. INTRODUCTION

While every cell in an individual is expected to have the exactly same DNA sequence (germline inheritable genetic raw material), genomic landscape heterogeneity is commonly observed in some specific cell types, e.g., immune T-cell, cancer cell, etc. [1, 2]. Moreover, recent studies show that somatic mutations in normal brain and skin are much more common than previous appreciated, suggesting that intercellular genomic heterogeneity is a rule instead of an exception [3-5]. Intercellular heterogeneity is a major confounding factor in studying individual populations that cannot be resolved directly by global profiling [6].

The complexity of heterogeneity has clinical implications. For example, a more heterogeneous tumor is more likely to fail therapy due to increased drug-resistant variants, and characteristics of dominant clones will not necessarily predict the behaviors of driver clones [7-9]. Intermingled intercellular genomic heterogeneity is often manifested by multiple clones with distinct sequences that cannot be resolved readily by global sequencing [10].

An experimental solution to mitigate intercellular heterogeneity is to isolate pure cell populations before sequencing; however, the methods are expensive, tedious, potentially biased and inapplicable to existing sequence data. On the other hand, thanks to the capacity of reading a genomic region multiple times, the next-generation sequencing (NGS) data potentially provides an alternative way to characterize genomic landscape heterogeneity. The major tasks at hand are to detect the genomic sequences of genuine clonal subtypes, and the number and abundance of clonal subtypes. Due to inevitable sequencing errors, systematic efforts to characterize intercellular genomic heterogeneity must effectively distinguish genuine clonal sequences from fake derivatives [11] (Fig. 1).

A quick but rough approach is to simply throw away the clonal subtypes whose read counts are less than a pre-specified threshold. Major limitations associated with this strategy are as the following. (1) The threshold is often arbitrary and hard to set. Indeed, different experiments may require different thresholds. It is also likely that different types of fake clones such as A->C or A->T may necessitate a separate threshold. (2) Clonal subtypes with higher read counts than the threshold can be fake, and similarly, lower read counts than the threshold can come from a genuine clone, resulting in both large false positives and large false negatives. (3) No quantitative quality measure can be assigned to the selected clonal subtypes (confidence level). Specifically, existing approaches largely ignore (a) significant differences in the misreading error rates on different nucleotide types (A, C, T, G) [11] and (2) differences in the "expected" read counts of fake clones caused by the parent genuine clones due to highly variable (depth) read counts across genuine clones.

Let us use the Fig 1 to both illustrate the limitation of existing method and suggest a viable solution. There are two ground-truth clones with sequences "ATGCTGCTGTGTA CTACTGCCTCGTGGGGC" and "ATGCTGCTGTGTACT ACTGCCTCGTGGGGC", respectively. The only difference is in the middle of the sequence. Clone 1 possesses "AC" while clone 2 possesses "CC". Actually, they have only 1 nucleotide difference. However, from the sequence data we can see there are totally five clones, which illustrates the point that the number of fake clones is often larger than the one of genuine clones. It is obvious that there is no good threshold available, because one genuine clone has three reads and at the same time there is a fake clone also having three reads. If the threshold is larger than three, one genuine clone will be missed. If the threshold is less than three, one fake clone will be falsely detected as genuine clone.

However, if we simultaneously explore the number of reads and the error rate, we may be able to reliably distinguish the two 3-reads clones. Suppose the error rates from A->G and A->C are 20% and 6%, respectively. Note that the error rates in real sequencing data are much lower than the ones we assume here, but we do observe that A->G error is three times more likely than A->C error. For the sake of discussion, we use the middle two nucleotides to represent each of the five clones. For example, clone "GC" is a fake clone and clone "CC" is a genuine clone, and both clones have three reads. According to the error rate we expect to see 12*0.2=2.4 reads of clone "GC" if we assume that the reads of clone "GC" is purely due to error. On the other hand, we expect to see 12*0.6=0.72 read of clone "CC". Compared to the real observation of 3 reads, we can conclude that clone "GC" is more likely due to error and hence fake, and clone "CC" is less likely due to error and hence genuine.

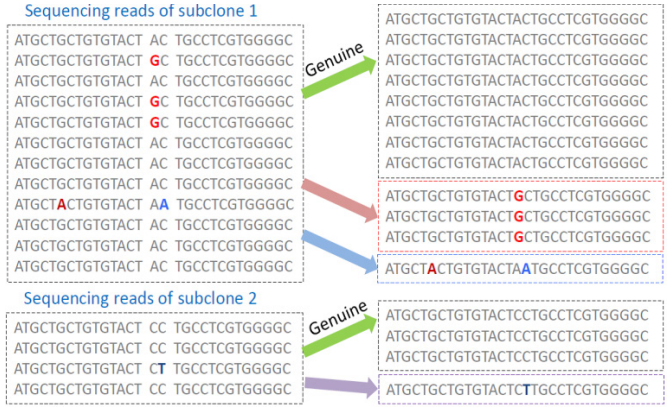

Figure 1. Illustration of genuine-fake mixed sequence clusters with variable sizes: raw sequence reads suggest five tumor subclones, while actually, there are only two genuine subclones.

In this paper we describe a statistical sequence modeling approach for detecting significant intercellular genomic heterogeneity (SIGH) directly from mixed sequencing reads rather than variant signatures. The approach was developed to identify the number and sequences of genuine subclones justified by model-based significance tests. In our model, the technical artifacts include but are not limited to the errors resulting from the sample preparation, PCR amplification, sequence reading and mapping. The approach is flexible to capture all kinds of artifacts and hence minimize the false positives.

SIGH basically works by exploiting the statistical differences in both the sequencing error rates at different nucleobases and the read counts of fake sequences in relation to genuine clones of variable abundance. We conduct a simulation study to demonstrate the performance of SIGH under a variety of conditions on realistic synthetic data.

To our best knowledge, the work that is closest to ours is [10], which applied the finite mixture model to estimate the tumor subtypes. However, it is essentially a sophisticated extension of the threshold method and largely ignored the fundamental relationship between clones embodied by various error rates and variable read counts across genuine clones that we have explored.

II. METHODS

A. Problem Formulation

Consider a tissue sample that exhibits intercellular heterogeneity. Assume the sample contains $M$ observed distinct clonal sequences that represent $M$ candidate subclones. Let $\beta_{ij}$ be the probability that the sequence of the $j$th candidate subclone is misclassified into the $i$th candidate subclone due to sequencing or other errors; $n_i$ be the observed number of reads in the $i$th candidate subclone; and $n_{ij}$ be the hidden number of misclassified reads in the $i$th candidate subclone that originate from the $j$th candidate subclone.

Hence, $n_{ij}$ follows binomial distribution $B(n_j + n_{ij}, \beta_{ij})$, where $n_j + n_{ij}$ can be interpreted as the number of total reads originating from the $j$th clone and $\beta_{ij}$ represents the error rate that the $j$th clone is misread as the $i$th clone. Based on the fact that in NGS data, $\beta_{ij}$ is usually quite small (smaller than 1%) for paired-end sequencing, the binomial distribution can be approximated quite well by Poisson distribution with parameter of $(n_j + n_{ij})\beta_{ij}$.

If we assume that the $i$th clone is fake, it must come from its parents. Observing that multiple different parental sequences can give birth to the same child sequence, we consider all other clones as potential parental clones. So, under the null hypothesis that the $i$th clone is fake, we have $n_i = \sum_{j=0, j \neq i}^{M} n_{ij}$. Because $n_{ij}$ is independent to each other for different $j$, $n_i$ follows Poisson distribution with parameter $\lambda_i$ defined by,

$$\lambda_i = \sum_{j=0, j \neq i}^{M}(n_j + n_{ij}) \beta_{ij} \approx \sum_{j=0, j \neq i}^{M} n_j \beta_{ij}, \quad (1)$$

where the last approximation comes from the expectation that the reads of fake clone are much smaller than its parental genuine clone, namely, $n_{ij} \ll n_j$.

Thus, under the null hypothesis that the $i$th clone is fake, if the number of reads associated with the $i$th clone is $k$, the distribution function is,

$$p(n_i = k) = \frac{\lambda_i^k e^{-\lambda_i}}{k!}, \text{ with } \lambda_i \approx \sum_{j \neq i} n_j \beta_{ij}, \quad (2)$$

i.e., the null distribution model solely due to errors.

B. Method Description

In the discussion above, we assumed the error rate $\beta_{ij}$ is known, however, we often don't know it *a priori*.

Fortunately, there are many scenarios under which the error rate can be reliably derived based on the estimate of more fundamental quantities. For example, immune cell sequencing covers both the somatic recombination region and constant region, where the somatic recombination region can be used to define clones and the constant region can be used to learn the error rates. Another example is the tumor cell sequencing. The availability of paired tumor-normal sequences makes it possible to determine which base is misread and hence the error probability [11].

Let $S$ be the true nucleobase; $R$ be the observed nucleobase; and $S$ or $R$ takes one of the four nucleobases (A, C, G, T), so there are 12 error types in total. We have observed that the error probability is not uniform and significantly depends on both genuine and erroneous nucleotide types. Hence, it is necessary to learn all 12 kinds of error probability. At a single base, the (misread) error probability can be estimated by $\Pr(R=r|S=s) = N_r/N_s$, where $N_r$ is the read counts with the observed nucleobase $r$ and $N_s$ is the read counts with the true nucleobase $s$. The estimate above illustrates the main idea, however, more care needs to be taken in real applications. Even in the constant region of the immune cells, polymorphisms may be confounded as errors if it is not recorded in the public database. A remedy trick is to disregard the seemingly extraordinary large error counts, because the polymorphism will appear as error of 100% for homozygous mutation and error of 50% for heterozygous mutation. Empirically, we found 10% is a good threshold since the real error is often less than 1%.

Thus, considering a clone is represented by a sequence containing multiple nucleotides, the probability of the $j$th candidate subclone is misclassified into the $i$th candidate subclone due to sequencing or other error is

$$\beta_{ij} = \prod_{l}^{L} \Pr\left(R = B_i(l) | S = B_j(l)\right), \quad (3)$$

where $B_i(l)$ is the nucleobase at the $l$th location in the sequence of the $i$th candidate subclone, L is the sequence length, and '$\beta_{ij} = 0$' if the $i$th and $j$th candidate subclones have different sequence lengths.

Once the error probabilities have been estimated and the number of reads supporting a clone is observed, we take the framework of hypothesis testing to help assess the probability of seeing such a clone under the null hypothesis. Given the observed read counts $n_i$ of the $i$th candidate subclone, the p-value under the null hypothesis (the $i$th candidate subclone is fake) is

$$p_i = \sum_{k=n_i}^{\infty} \frac{\lambda_i^k e^{-\lambda_i}}{k!}. \quad (4)$$

Accordingly, at the significance threshold $T_\alpha$, the $i$th candidate subclone is considered as genuine if $p_i \leq T_\alpha$; otherwise, it is fake. The rationale behind is simple. If the clone is unlikely to happen under the hypothesis of fake clone, that is, the p-value is smaller than the threshold, it should be considered as a genuine clone.

SIGH determines $T_\alpha$ by controlling the false discovery rate, which is defined as the expected proportion of false positives among all significant hypotheses [12], via: (1) sort the p-values in an increasing order denoted by p(1),…, p(M); (2) find the largest $m$ such that $p(m) \leq m\alpha / M$; (3) set $T_\alpha = p(m)$ for the significance level $\alpha$. The reliability of the detected genuine subclones is assured by the quantitative significance measure.

### III. SIMULATION STUDIES AND RESULTS

To demonstrate the performance of SIGH under various conditions, we conducted simulation studies based on real sequencing data from our in-house immune T-cells dataset. Our simulations adopt settings similar to [10], with $L$=100. We first simulated sequence read data for normal samples by randomly replacing the nucleobase $s$ by $r$ at each locus of the baseline sequence, according to an error probability table (Table 1).

TABLE I. PARAMETER SETTINGS AND RESULTS. THE NUMBERS OF OBSERVED AND GENUINE CLONAL SEQUENCE CLUSTERS IDENTIFIED BY SIGH ARE REPORTED.

| Probability ($10^{-4}$ except diagonal) | | | | Case | # of clones | | |
|---|---|---|---|---|---|---|---|
| | A | C | G | T | # | $M$ | $M_0$ | $M_{SIGH}$ |
| A | 0.99 | 0.29 | 8.70 | 0.48 | 1 | 171~250 | 3 | 3 |
| C | 0.27 | 0.99 | 0.38 | 9.62 | 2 | 180~423 | 5 | 5 |
| G | 4.82 | 0.19 | 0.99 | 0.45 | 3 | 257~625 | 10 | 10 |
| T | 0.62 | 7.48 | 0.32 | 0.99 | *All detections are made at $p$=0.05. | | | |

We ran similar simulations with $M_0$ = 3, 5, 10 genuine subclones of distinct sequences for various clonal proportions to determine how well SIGH could identify genuine subclone sequences and characterize intratumor heterogeneity. The distributions of proportions of genuine clones are varying across the 15 datasets as shown in Table 2. Since error probability is generally much smaller than that of random polymorphisms, only the loci with error rates smaller than 10% were used in the estimation. We tested SIGH on 15 datasets each with unique set of conditions. For each application of SIGH, mean per-sample execution time was ~2 minutes on a computer equipped with an Intel® Xeon ® CPU X5660 @ 2.80GHz (24 cores and 23.5 GB of RAM).

TABLE 2. THE PROPORTIONS OF EACH CLONE IN EACH DATASET AND THE DETECTED RESULTS OF GENUINE CLONES FROM SIGH

| Data set | Proportions of each genuine clone | SIGH (all clones, genuine clones) |
|---|---|---|
| 1 | (0.33,0.33,0.33) | (250, 3) |
| 2 | (0.90,0.09,0.01) | (177, 3) |
| 3 | (0.70,0.20,0.10) | (195, 3) |
| 4 | (0.831,0.166,0.003) | (172, 3) |
| 5 | (0.01,0.98,0.01) | (171, 3) |
| 6 | (0.20, 0.20, 0.20, 0.20, 0.20) | (423, 5) |
| 7 | (0.009,0.003,0.006,0.893,0.089) | (180, 5) |
| 8 | (0.71,0.17,0.02,0.09,0.01) | (254, 5) |
| 9 | (0.002,0.395,0.592,0.001,0.010) | (181, 5) |
| 10 | (0.61,0.20,0.08,0.10,0.01) | (283, 5) |
| 11 | (0.10,0.10,0.10,0.10,0.10,0.10,0.10,0.10,0.10,0.10) | (625, 10) |

| 12 | (0.167,0.001,0.033,0.668,0.067, 0.010,0.027,0.017,0.007,0.003) | (257, 10) |
| 13 | (0.208,0.001,0.104,0.416,0.104, 0.021,0.062,0.062,0.011,0.011) | (395, 10) |
| 14 | (0.152,0.001,0.061,0.606,0.121, 0.009,0.030,0.015,0.003,0.002) | (283, 10) |
| 15 | (0.116,0.001,0.047,0.698,0.093, 0.007,0.023,0.012,0.002,0.001) | (279, 10) |

For the normal sample simulations, SIGH accurately learned the reading error probabilities in all datasets, with small estimation errors against the ground truth. The statistics (boxplot: green stars indicate the ground truths) on the estimated error probabilities across the 15 datasets are summarized in Figure 2.

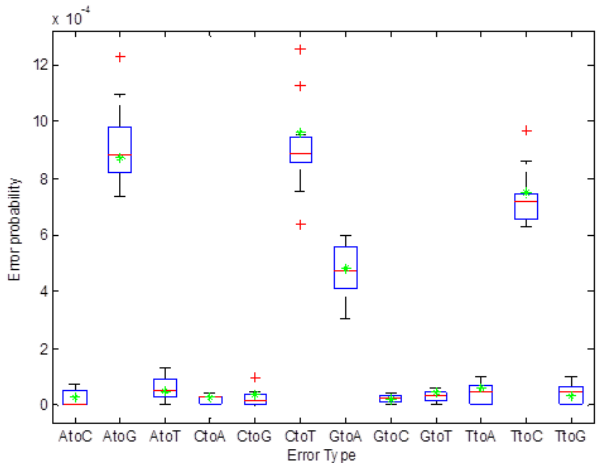

Figure 2. Summary of error probabilities estimated by SIGH in 15 simulation datasets; green stars are ground truth from Table 1.

For the heterogeneous tissue with variable subclonal proportions, SIGH correctly identified the number of genuine subclones from an overwhelmingly large number of fake sequences, in all 15 datasets, reported in Tables 1&2. As a significance testing approach, measures of confidence are directly derived from the null distribution of $n_i$ in the form of p values (Table 1). SIGH also accurately estimated the proportions of genuine clonal subpopulations. As expected, SIGH works well on data with balanced clone population, that is, all clone proportions are similar. It is worth noting that SIGN also works well on very imbalanced data. For example, in data set 15, the most common clone has proportion of 69.8%, while the rarest clone occupies merely 0.1%, a difference of 700 folds. The capability of SIGN for imbalanced data may be particularly useful for the detection of rare clones.

## IV. DISCUSSIONS AND CONCLUSIONS

Our realistic simulation studies demonstrate the feasibility of identifying intercellular genomic heterogeneity in the presence of inevitable and variable sequencing errors. By statistically modeling read count null distributions for each of the observed sequence clusters, SIGH can potentially identify rare genuine subclone(s) directly from genuine-fake mixed raw or mutation sequence reads, despite abundance disparity between subclones. When applied to longitudinal studies, SIGH may also provide additional information about tumor evolution or clonal repopulation dynamics [8].

We foresee a variety of extensions to the concepts in SIGH. For example, with further development, SIGH methodology can be applicable to analyzing the raw sequences of enriched immune cells in tumors. The potential ability to detect sequence reads associated with somatic immune cells is clinically significant, since somatic immune cells play a critical role in tumor micro-environment heterogeneity [13]. SIGH analysis on somatic T-cells may reveal migration of immune cells (recruitment and localization) and immune suppression. In fact, immune infiltration can include multiple cell types, having both pro- and anti-tumor functions with varying activation statuses and localizations within the tumor [13].

*Funding*: National Institutes of Health, under Contract HHS-N2612200800001E, Grants HL111362 and CA160036, in part.

*Conflict of Interest*: none declared.